\documentstyle[12pt]{article}
\begin{document}

\baselineskip=20pt

\begin{titlepage}
\begin{center}
{\Large\bf
Dilepton production from resonance scattering \\in hot hadronic matter}\\
\vspace{0.2in}
Chungsik Song and C. M. Ko\\
\vspace{0.2in}
{\small\it Cyclotron Institute and Physics Department\\
Texas A\&M University, College Station, TX 77843, USA }\\
\vspace{0.1in}
(July, 1995)
\vspace{0.2in}

{\bf Abstract}
\end{center}
Dilepton production from resonance scattering in hot hadronic matter
is studied. Including the widths of these resonances, which enhance the
phase space for dilepton production,
we find that the production rate is significantly increased
if a resonance appears in the extended phase space.
For the reaction $\pi+\rho\to l^++l^-$, the finite $\rho$ meson width
extends the invariant mass below the $\omega$-meson mass, so a peak
at the $\omega$ meson mass is seen in the dilepton spectrum. Similarly,
a $\rho$-meson peak appears in the reaction $\rho+\rho\to l^++l^-$.
On the other hand, the effect
of particle widths in the reaction $\pi+a_1\to l^++l^-$ is
small since the extended phase space does not include any resonance.

\end{titlepage}

\newpage

\section{Introduction}

At very high temperature and/or density, as achieved in ultrarelativistic
heavy ion collisions \cite{rhic}, the hadronic matter is expected to undergo
a phase transition to the quark-gluon plasma
in which quarks and gluons interact weakly and the spontaneously broken
chiral symmetry is restored \cite{qgp}.
One possible way to
verify the existence of the quark-gluon plasma in these
collisions is to detect the produced photons and dileptons.
Since these electromagnetically interacting particles have mean free paths
much larger than the transverse size of the colliding system,
they leave the system without
further interactions and thus carry information about the hot dense matter
in which they are produced \cite{dilepton,larry}.

However, photons and dileptons can also be produced
from hadronic matter below the phase transition temperature and density.
To find the signals for the quark-gluon plasma needs thus a good
understanding of this background contribution. On the other hand, dilepton
production from hadronic matter is useful for
understanding precursor phenomena, such as the change of meson masses and
widths, related to chiral symmetry restoration
and the deconfinement phase transition in hot dense
hadronic matter \cite{pisaski,kapusta1,ko,hkl}.

Recently, photon \cite{kapusta2,xsb,song1,kevin1}
and dilepton \cite{lichard,song2}
production from hadronic matter have been
extensively studied with the effective Lagrangian.
It has been shown that
in hadronic matter reactions with $a_1$ meson in the intermediate state,
e.g., $\pi+\rho\to a_1\to \pi+\gamma$,
significantly enhance the photon production rate.
Also, for dileptons with invariant masses in the
region 1.5 $\sim$ 3.0 MeV,
which has been suggested as a possible window for
observing dileptons from the quark-gluon plasma,
contributions from resonance scattering,
such as $\pi+\rho\to l^++l^-$, $\rho+\rho\to l^++l^-$, $\pi+a_1\to l^++l^-$,
and etc., are important.
Recent experimental data from CERN/SPS
experiments seem to be consistent with these observations
\cite{photon,dilep}.

In previous studies, some important features of
hadron in-medium properties have not been included.
The present authors and Lee \cite{slee}
have shown that the pion electromagnetic
form factor is reduced at finite temperature, leading thus to a
suppression of dilepton production from pion-pion annihilation.
The reduction of the from factor at finite temperature, which can be related
to the deconfinement transition
and chiral symmetry restoration in hot hadronic matter,
has also been shown in calculations based on the
QCD sum rule at finite temperature \cite{domi} and
the QCD factorization
assumption at large momentum transfer \cite{satz}.

Another feature, which is the subject of present paper, is that most
mesons are not stable and easily decay into other mesons.
The widths of these mesons are not
negligible even in free space, e.g., $\Gamma_\rho\sim$ 150 MeV and
$\Gamma_{a_1}\sim$ 400 MeV for the $\rho$ and $a_1$ mesons, respectively,
and, furthermore, they are also expected to change with temperature.
The finite particle width increases the phase space for dilepton production
and thus modifies its production rate from the hadronic matter.
In this letter, we shall calculate the dilepton production rate
from hot hadronic matter by
including the particle widths.  This study can be easily
extended to photon production from hot matter.

\section{Formalism}

We consider dilepton production from a reaction in which two
hadrons ($h_1,\>h_2$) annihilate into a pair of leptons
through a resonance ($r^*$), i.e.,
$h_1+h_2\to r^*\to l^++l^-$, as shown in Fig.~1.
The dilepton production rate $(R)$/volume$(V)$ from
such a reaction in hot matter can be
written as \cite{larry}
\begin{eqnarray}
{R\over V} &=& -4e^4\int{d^3{\bf k}_1\over (2\pi)^3 2\omega_1}
                        {d^3{\bf k}_2\over (2\pi)^3 2\omega_2}
                        {d^3{\bf p}_1\over (2\pi)^3 2E_1}
                        {d^3{\bf p}_2\over (2\pi)^3 2E_2}
               (2\pi)^4\delta^{(4)}(k_1+k_2-p_1-p_2)\cr
          & &\qquad\times f(\omega_1) f(\omega_2)l^{\mu\nu}(p_1,p_2)
             \left({1\over q^2}\right)^2 h_{\mu\nu}(k_1,k_2).
\end{eqnarray}
In the above $k_i=(\omega_i,{\bf k}_i)$ and $p_i=(E_i,{\bf p}_i)$
are, respectively, the four momenta of the colliding hadrons and dileptons,
and $q=p_1-p_2$;
$f(\omega)$ is the Fermi-Dirac or Bose-Einstein distribution function;
$h_{\mu\nu}(k_1,k_2)$ is the hadronic tensor which
depends on the reaction;
and the leptonic tensor $l^{\mu\nu}(p_1,p_2)$ is given by
\begin{equation}
l^{\mu\nu}(p_1,p_2)=(p_1\cdot p_2+m_l^2)g^{\mu\nu}
                    -p_1^\mu p_2^\nu-p_1^\nu p_2^\mu.
\end{equation}

In eq.~(1) it has been assumed that the colliding particles are on
mass shell, i.e., $\omega_i=\sqrt{{\bf k}_i^2+m_i^2}$. However,
these particles in general have finite widths and should
be described by spectral functions
$\rho_i(\omega_i,{\bf k}_i)$. This requires the following modification
\cite{kox}
\begin{equation}
\int{d^3{\bf k}\over (2\pi)^32\omega}\rightarrow\int{d^3{\bf k}\over (2\pi)^3}
\int{d\omega^\prime\over (2\pi)}\rho(\omega^\prime,{\bf k}),
\end{equation}
and leads to
\begin{eqnarray}
{d(R/V)\over dM^2} &=& -4e^4\int{d^4k_1\over (2\pi)^4}
                              {d^4k_2\over (2\pi)^4}
               \rho_1(\omega_1,{\bf k}_1)\rho_2(\omega_2,{\bf k}_2)\cr
               & & \qquad\times f(\omega_1) f(\omega_2)
                h_{\mu\nu}(k_1,k_2)L^{\mu\nu}(k_1,k_2,M^2),
\end{eqnarray}
where $M$ is the invariant mass of the dilepton, $M^2=(p_1+p_2)^2$, and
$L_{\mu\nu}(k_1,k_2,M^2)$ is given by
\begin{eqnarray}
L_{\mu\nu}(k_1,k_2,M^2) &=& \int {d^3{\bf p}_1\over (2\pi)^3 2E_1}
                            \int {d^3{\bf p}_2\over (2\pi)^3 2E_2}
(2\pi)^4\delta^{(4)}(k_1+k_2-p_1-p_2)\cr
& &\qquad\times\delta[(p_1+p_2)^2-M^2]
l_{\mu\nu}(p_1,p_2)\left({1\over q^2}\right)^2.
\end{eqnarray}
For dileptons at rest in the center-of-momentum frame,
this can be written as
\begin{equation}
{dL^{\mu\nu}\over d^3{\bf q}}\Biggl\vert_{{\bf q}=0}
={1\over 24(2\pi)M^3}\delta(K_0-M)\delta^{(3)}({\bf K})
\left(g^{\mu\nu}-{Q^\mu Q^\nu\over M^2}\right),
\end{equation}
with $Q=(M,0,0,0)$ and
$(K_0,{\bf K})=(\omega_1+\omega_2,\,{\bf k}_1+{\bf k}_2)$.

{}From eqs.~(4) and (6), we have
\begin{eqnarray}
{d(R/V)\over dMd^3{\bf q}}\Biggr\vert_{{\bf q}=0}
&=&{e^4\over 6(2\pi)^9M^3}
 \int d^3{\bf k}_1dw_1 \rho_1(\omega_1,{\bf k}_1)\rho_2(M-\omega_1,
 -{\bf k}_1)\cr
 & &\qquad\times f_1(\omega_1)f_2(M-\omega_1)W(\omega_1,{\bf k}_1,M),
\end{eqnarray}
with
\begin{equation}
W(\omega_1,{\bf k}_1,M)=h^{\mu\nu}(k_1,Q-k_1)
\left(-g_{\mu\nu}+{Q_\mu Q_\nu \over M^2}\right).
\end{equation}
One can easily check that when particles are taken to be on mass shell, i.e.,
$\rho(\omega,{\bf k})=2\pi\delta(k^2-m^2)$, eq. (7) reduces to the
one used in the literature \cite{kapusta1}.

\section{Effective Chiral Lagrangian}

We assume that the hadronic matter consists of pions,
vector mesons, and axial vector mesons and that their interactions are
described by an effective chiral Lagrangian.
The interactions among these hadrons are constrained by
symmetry properties of the strong interaction.
The resulting parameters in the Lagrangian
are then determined from the experimental data on hadron
properties \cite{weinberg}.
The chiral Lagrangian for pseudoscalar mesons at low
energies takes the form
\begin{equation}
{\cal L}=-{1\over8}F_\pi^2{\rm Tr}(\partial_\mu U\partial^\mu U^\dagger),
\end{equation}
where $F_\pi=135$MeV is the pion decay constant,
and $U$ is related to pseudoscalar mesons $P$ by
\begin{equation}
U={\rm exp}\left[{2i\over F_\pi}P\right],
\qquad P=\pi^a{\lambda^a\over\sqrt{2}}.
\end{equation}
Vector ($V_\mu$) and axial vector ($A_\mu$) mesons are
included as massive Yang-Mills fields of the chiral symmetry \cite{vmd}
and the derivative in eq.~(9) is thus replaced by the covariant one, i.e.,
\begin{equation}
\partial_\mu U\rightarrow(\partial_\mu U-igA_\mu^L U+igUA_\mu^R).
\end{equation}
where $A_L^\mu=(V^\mu-A^\mu)/2$ and $A_R^\mu=(V^\mu+A^\mu)/2$.
The phenomenological gauge coupling constant ($g$)
can be determined by fitting the masses and widths
of vector and axial vector mesons \cite{song3}.
The electromagnetic interaction is then introduced through the notion of
vector meson dominance (VMD) \cite{sakurai}, i.e.,
\begin{equation}
{\cal L}_{em}=-{\sqrt{2}e\over g}a^\mu\left[m_\rho^2\rho^0_\mu
+{1\over3}m_\omega^2\omega_\mu-{\sqrt{2}\over3}m_\phi^2\phi_\mu\right].
\end{equation}

We also include the gauged Wess-Zumino term
in the effective Lagrangian to describe the non-Abelian anomaly structure
of QCD \cite{witten}, which
leads to an anomalous interaction among a pseudoscalar meson and two
vector mesons,
\begin{equation}
{\cal L}_{VVP}=-{3g^2\over16\pi^2F_\pi}\epsilon^{\mu\nu\alpha\beta}
                   {\rm Tr}(\partial_\mu V_\nu\partial_\alpha V_\beta P),
\label{vvp}
\end{equation}
where $\epsilon^{\mu\nu\alpha\beta}$ is the antisymmetric Levi-Civita tensor
with $\epsilon^{0123}=1$.

{}From eq.~(\ref{vvp}) we have
\begin{equation}
{\cal L}_{\omega\pi\rho}=-\sqrt{2}g_\omega\epsilon^{\mu\nu\alpha\beta}
                         \partial_\mu \omega_\nu
                         \partial_\alpha\rho_\beta\cdot\pi,
\end{equation}
with
\begin{equation}
g_\omega=\left({3g^2\over 16\pi^2F_\pi}\right).
\end{equation}
Combining with the vector meson dominance assumption, we obtain
\begin{equation}
\Gamma(\omega\to\pi^0\gamma)={3\over64\pi^4}{\alpha g^2\over F_\pi^2}
\vert{\bf q}_\pi\vert^3=0.80 {\rm MeV}.
\end{equation}
This is in good agreement with the experimental value,
$\Gamma_{exp}(\omega\to\pi^0\gamma)=0.86\pm 0.05$ MeV.
There is another anomalous decay, $\phi\to 3\pi$,
which violates the OZI rule.
We regard this decay as being due to a small $\omega-\phi$ mixing:
\begin{eqnarray}
\omega_\mu=\tilde\omega_\mu-\epsilon\tilde\phi_\mu,\cr
\omega_\mu=\epsilon\tilde\omega_\mu+\tilde\phi_\mu,
\end{eqnarray}
where a tilde denotes physical fields.
{}From the anomalous $VVP$ interaction, we find
\begin{equation}
\Gamma(\phi\to\rho\pi)={9g^4\vert{\bf q}_\pi\vert^3\over512\pi^5 F_\pi^2}
                       \epsilon^2,
\end{equation}
where ${\bf q}_\pi$ is the pion three-momentum in the rest frame
of the $\phi$ meson.
{}From the experimental value $\Gamma_{exp}(\phi\to\rho\pi)\approx 0.62$MeV,
we get
$\vert\epsilon\vert=0.076$, which is in fair agreement with
that obtained from the canonical-mass-mixing model, $\epsilon=-0.058$
\cite{phi}.

\section {Results and Discussions}

We first consider dilepton production from
$\pi+\rho$ annihilation based on the vector meson dominance.
The hadronic tensor is given by
\begin{equation}
h_{\mu\nu}=2\biggl\{[q^2 k^2-(q\cdot k)^2]g_{\mu\nu}
+q\cdot k (q_\mu k_\nu+q_\nu k_\mu)-k^2q_\mu q_\nu-q^2 k_\mu k_\nu\biggr\}
\vert F_{\pi\rho}(q)\vert^2,
\end{equation}
where $k$ and $q$ are the four momenta of the $\rho$ meson and the intermediate
vector meson, respectively.
The form factor $F_{\pi\rho}(q)$ is obtained from the VMD assumption
and is given by
\begin{equation}
\vert F_{\pi\rho}(q)\vert^2=
\vert F_\omega(q)+F_\phi(q)\vert^2.
\end{equation}
In terms of the photon-vector meson coupling constant,
$g_{{\rm v}\gamma}$, and the vector meson width $\Gamma_V$,
$F_V(q)$ is written as
\begin{equation}
F_V(q)={g_{\rm v}g_{{\rm v}\gamma}\over q^2-m_V^2-im_V\Gamma_V},
\end{equation}
with $g_\phi=\epsilon g_\omega$.

{}From Eq.~(1), we can then easily obtain the dilepton production rate for
the case without including widths of the colliding particles.
Since both $\omega$ and $\phi$ mesons can appear as intermediate states of
the reaction $\pi+\rho\to l^++l^-$, their interference needs to be included.
The result at $T=$ 150 MeV is shown by the dotted curve in Fig.~2.
If the interference effect is neglected, then the result
is similar to that obtained in Ref.~\cite{lichard}.
In the calculation, we have neglected the lepton masses
but it is straightforward to include them.

To include particle widths, we assume that
the $\rho$ meson spectral function, defined via
\begin{equation}
\langle 0\vert V^\mu_a(x) V^\nu_b(0)\vert 0\rangle =
-{\delta_{ab}\over (2\pi)^3}\int d^4k\theta(k^0)e^{ik\cdot x}
\left(g^{\mu\nu}-{k^\mu k^\nu\over k^2}\right) \rho_V(k),
\end{equation}
has a Breit-Wigner form,
\begin{equation}
\rho_V(k) = \frac{2m_\rho\Gamma_\rho}
              {(k^2-m_\rho^2+\Gamma^2_\rho/4)^2+m_\rho^2\Gamma_\rho^2},
\end{equation}
with $\Gamma_\rho=150$ MeV being the $\rho$ meson width in free space.
It becomes a delta function in the limit $\Gamma_\rho\to 0$, i.e,
\begin{equation}
\frac{2m_\rho\Gamma_\rho}
              {(k^2-m_\rho^2+\Gamma^2_\rho/4)^2+m_\rho^2\Gamma_\rho^2}
\rightarrow 2\pi\delta(k^2-m_\rho^2).
\end{equation}
For a pion, we assume that
\begin{equation}
\rho_\pi=2\pi\delta(k^2-m_\pi^2).
\end{equation}

The dilepton production rate is then calculated from eq.~(7).
Since the pion spectral function is given by a delta
function we are left with only an one-dimensional integral which can
be easily evaluated numerically. The result obtained with the inclusion of
the rho meson width is shown in Fig.~2 by the solid curve.
We see that there is now
a large enhancement in the dilepton production rate
near the $\omega$ resonance as a result of the extension of
the phase space integral.
Without including the $\rho$ meson width, it is not possible
to have the $\omega$ meson peak
in the dilepton spectrum from the $\pi+\rho$ reaction since the threshold
value for the invariant mass is above the $\omega$ meson mass.
This result is compared with that from the $\pi+\pi$ reaction through
the $\rho$ resonance shown by the dashed curve in Fig. 2.
We find that the $\omega$ meson peak is above the $\rho$ meson peak from
the $\pi+\pi$ reaction due to the very narrow
$\omega$ meson width, $\Gamma_\omega\sim$ 8.5 MeV.
The interference between the $\omega$ and $\phi$ meson intermediate states
leads to a reduction in the
production rate when $M\ge m_\phi$ and
is most appreciable at $M\sim m_\phi$ as a result of $g_\omega\gg g_\phi$.
However, the effect cannot be observed as it
is far below the contribution from the $\pi+\pi$ reaction.

For dileptons from the $\rho+\rho$ reaction, the result at $T=150$ MeV
including the particle widths is shown in Fig.~3 by the solid curve.
We see in the spectrum a $\rho$ peak which is absent in the case
without including the particle widths as shown by the dotted curve.
However, its magnitude near the
resonance is smaller than that from $\pi+\pi$ annihilation
(the dashed curve) as a result of the large $\rho$ meson width.
In Fig.~4, the result for the $\pi+a_1$ reaction including the
$a_1$ meson width is shown by the solid curve.
It is seen that the $a_1$ meson width has only a small effect
on the $\pi+a_1$ reaction
as no new resonance appears in the
extended phase space when particle widths are included.
Compared with the case without including
particle widths shown by the dotted
curve, there is a small reduction of the production rate near the threshold
and also an extension of the production rate into
the low invariant mass region.

In hot hadronic matter, mesons change their properties when
the temperature of the system
approaches the critical value for chiral
symmetry restoration and the deconfinement phase transition.
In particular, meson widths will be modified by collisions in hot matter.
We consider the $\pi+\rho$ reaction in which
a sharp peak near the $\omega$ meson resonance appears
as a result of its narrow width in free space.
Effects due to in-medium widths of the colliding particles
are expected to be small because no new resonance is involved.
However, the broadening of the resonance in the
intermediate state will affect appreciably the dilepton production rate.
The dominant contribution to the $\omega$ meson width comes from
scattering processes in which the $\omega$ meson interacts with thermal
pions through the $b_1(1236)$ resonance \cite{kevin2}.
To estimate the broadening of its width in hot matter,
we use the relation,
\begin{equation}
\Gamma^{coll}_\omega=\int{dp^3_\pi\over (2\pi)^3}
f_\pi(E_\pi)\sigma_{\pi\omega}(M)v_{rel},
\label{width}
\end{equation}
and assume a Breit-Wigner form for the cross section,
\begin{equation}
\sigma_{\pi\omega}(M)={\cal N}{\pi\over \vert{\bf k}\vert^2}
{\Gamma^2_{b_1\to\pi\omega}\over (M-m_{b_1})^2+\Gamma^2_{b_1}/4},
\end{equation}
where $\cal N$ is the degeneracy factor from the spin and isospin of
the colliding particles and resonances.
The $\omega$ meson width is found to increase with temperature, and
$\Gamma^{coll}\sim$ 12 MeV at $T=150$ MeV \cite{diff}.

The effect due to  collisional broadening of the intermediate resonance
on the $\pi+\rho$ interaction is shown in Fig.~5 by the solid curve.
We also include the changes in the $\rho$ and $\phi$ meson widths with
temperature, which
is estimated to be $\Gamma_{coll}\sim$ 25 MeV \cite{rho} and 10 MeV \cite{ks}
at $T=150$ MeV, respectively.
We see that the effect of a modified $\omega$ meson width at
finite temperature smooths out the peak in the dilepton spectrum.
However, there is still a large enhancement of dileptons with invariant
masses around the $\omega$ meson when the width of rho meson is included.

\section{Conclusion}

Dilepton production from resonance scattering
at finite temperature is studied by including widths
of the colliding particles. The finite widths
increase the phase space for dilepton production and
makes it possible that a resonance peak appears below the threshold
in the case without particle widths.
This  effect is significant
for $\pi+\rho$ and $\rho+\rho$ reactions
in which the extended threshold includes other resonances.
However, it has only small effects if there is no resonance appearing in the
extended phase space as shown in the $\pi+a_1$ reaction. In this case,
the production rate is slightly reduced near the threshold and is
further extended into lower invariant masses.
For the $\pi+\rho$ reaction, the interference between
the two possible reactions, $\pi+\rho\to\omega\to l^++l^-$ and
$\pi+\rho\to\phi\to l^++l^-$, leads to a reduction in
the dilepton production rate for $M\ge m_\phi$. However, the effect cannot be
observed because of the larger contribution from the $\pi+\pi$ reaction.

Particle widths are modified in hot matter, and this
also affects the dilepton production rate.   Again, the effect due to
changes in the widths of the colliding particles is small
unless new resonances appear in the extended
phase space. When thermal effects on the widths of intermediate resonances are
included, we obtain a reduction of the resonance peak in the
dilepton spectrum as shown in the $\pi+\rho$ reaction.
However, there is still a significant enhancement
of dileptons with invariant masses near the vector meson
resonance compared with the case
without including the widths of the colliding particles.

In future studies of dilepton production from nucleus-nucleus collisions,
one should thus include both finite particle widths and
the medium modification of hadron properties.  This is
necessary for identifying
the signatures of the quark-gluon plasma and
the precursor phenomena related to the phase transition from the
hadronic background. We expect that for photon production from
hot dense matter there will be similar effects, and we plan to investigate
this in the future.

\bigskip
\centerline{{\bf Acknowledgment}}

This work was supported in part by the National Science Foundation
under Grant No. PHY-9212209 and the Welch Foundation under Grant No. A-1110.
The support of CMK by a Humboldt Research Award is also gratefully
acknowledged, and he would like to thank Ulrich Mosel of the University of
Giessen for the warm hospitality.

\newpage

\centerline{{\bf Figure Captions}}

\vspace{1.5cm}

\noindent
{\bf Fig.\ 1}
Dilepton production from hadron scattering.
\vspace{1cm}

\noindent
{\bf Fig.\ 2}
The dilepton production rate from hot hadronic matter at $T$= 150 MeV.
The solid and dotted curves are, respectively, contributions
from $\pi+\rho\to l^++l^-$ with and without $\rho$ meson width, while the
dashed curve is from $\pi+\pi\to l^++l^-$.
\vspace{1cm}

\noindent
{\bf Fig.\ 3}
Same as Fig. 2 for $\rho+\rho\to l^++l^-$.
\vspace{1cm}

\noindent
{\bf Fig.\ 4}
Same as Fig. 2 for $\pi+ a_1 \to l^++l^-$.
\vspace{1cm}

\noindent
{\bf Fig.\ 3}
Same as Fig. 2 with in-medium $\rho$, $\phi$, and $\omega$ meson widths.


\begin{thebibliography}{40}
\bibitem{diff} The collision rate for $\omega$ meson is reduced by a factor
               of four compared with that
               in Ref.~\cite{kevin1} as we have used $\Gamma_{\pi+\omega\to
b_1}
               ={1\over2}\Gamma_{total}$.
\bibitem{rho} The collision rate for the $\rho$ meson is estimated in
              the same way as for the $\omega$ meson. We assume that
              the dominant
              contribution comes from the $a_1$ resonance.
              More detailed calculations show that the $\rho$ meson
              width increases by about 40 MeV at $T=150$ MeV \cite{kevin2}.
\end{thebibliography}
\end{document}